\documentstyle [12pt] {article}
\def\ni{\noindent}
\def\deg{\ifmmode^\circ\else$^\circ$\fi}

\begin{document}
\centerline{\Large \bf Grand Unified Model of Accretion Disks:}
\centerline{\Large \bf The Sub-Keplerian Paradigm$^a$}
\medskip
\centerline{SANDIP K. CHAKRABARTI$^b$}
\centerline{NASA/GSFC, Greenbelt, MD 20771}
\footnote{$^a$Supported in part by NRC Senior Research Associateship of
National Academy of Science}
\footnote{$^b$On leave from Tata Institute of Fundamental Research, Bombay,
400005, INDIA}

\baselineskip 14pt

In the 1970s, the standard accretion disk models were constructed$^{1,2}$
to largely explain observations from the binary systems. In these systems,
angular momentum supplied at the outer edge is necessarily Keplerian.
Viscosity drives the inflow by removing angular momentum outwards
and keeping the entire disk Keplerian in the process. Since then these
binary disk models are being used also to explain big blue bump observed
in the continuum of active galaxies and quasars and with some
success$^3$. Occasionally, one invokes additional components
along with the standard thin disks, such as corona, warm absorbers, etc.
in order to explain continuum as well as variable components of X-rays and
$\gamma$-rays. There are several recent observations of almost
zero time-lag correlated variabilities between X-rays and optical$^{4}$ which
cannot be explained by simple Keplerian disk models. Temporal variation
of line profiles from objects, such as, ARP102B and 3C390.3 is impossible
to explain by using axisymmetric disk models$^{5}$. Recent HST observation
of M87 shows clear evidence of non-axisymmetry in the ionized disk
around the black hole and the association of the spiral shock in NGC4258
with the accretion disk cannot be ruled out. These examples suggest the
existence of large scale spiral shock waves in both M87 and NGC 4258$^{6}$.
It is also generally believed that the radiation from our own galactic
center is most likely coming from a low efficiency, quasi-spherical, accretion
flow.

That the disks need not be of `standard' type was sensed by theoreticians
even in late '70s and throughout the '80s. Thick accretion disk,
transonic accretion disk and slim accretion disk models (For a
general discussion and references, see Chakrabarti$^7$) came
about. In thick disks, angular momentum is assumed to be almost constant
but the radial motion is ignored. In transonic disks, the flow is
thin but the radial motion is included. In slim disks, matter is allowed to
pass through the inner sonic point, thus improving on the standard
disk model. However, in all these models, unsuccessful attempts were
made to match the flow with {\it Keplerian disks} at some distance and the disk
structure depended strongly upon the matching radius and other parameters
invoked.

This intrinsic problem with these theoretical models, as well as problems
in explaining a large number observations with Keplerian disks, particularly
when applied to active galaxies, disappear with the realization
that the accretion disks in active galaxies need not be Keplerian anywhere
including the outer boundary! The outer boundary condition here is
completely different from that of the binary systems, since matter
is largely supplied by winds from mass lossing and colliding stars
in random motion. Matter could lose most of its angular
momentum before it brings itself together to form an accretion disk.

After the matter starts with highly sub-Keplerian angular momentum, its
subsequent behaviour depends strongly upon the accretion rate and the
viscosity in the flow. If the accretion rate is small enough, the flow passes
through the outer sonic point (just as a Bondi flow) and remains supersonic
before falling onto a black hole. If the accretion rate is high,
matter would pass through the inner sonic
point$^{8-9}$. This description is valid if the entropy remains
almost constant, i.e., the viscosity is small enough.
Even when the flow starts with small entropy, some entropy could be
generated at a shock or in the flow by viscosity which then
allows the flow to pass through the inner sonic point as well$^{10-12}$.
If the entropy is higher (for a given accretion rate) it is likely
that strong winds may be formed from flows with positive energy$^{9,11}$.
Except when the viscosity is high, the flow radiates with a very low
efficiency as in a Bondi flow as discussed in these works.

{\parfillskip=0pt Contrary to a Newtonian star, a black hole has no hard
surface. If the flow is unable to lose angular momentum efficiently, the
centrifugal barrier causes the flow to have a shock close to the black
hole$^{7-12}$. The postshock flow is the boundary layer equivalent
of a black hole accretion. This is clearly the case when the viscosity is
small enough. The postshock flow has all the features of a
thick accretion disk only more self-consistent since the
radial motion is also included$^{11}$. The preshock flow is mainly advected
towards the black hole, just as a Bondi flow, but the immediate post-shock
flow is rotationally dominated as in a Keplerian disk. Further on,
the flow picks up radial motion and supersonically enters into the
black hole. If the viscosity is high (typically, if the
$\alpha$ parameter$^1$ is larger than about $10^{-2}$) the {\it stable} shock
disappears and the disk eventually becomes Keplerian$^{10}$ except
close to the boundaries. \par }

\vspace{6.10cm}

{\noindent \small Fig. 1(a-b): Mach Number (Y-axis) as a function of
logarithmic
radial distance (X-axis) for viscous isothermal flows. In (a),
a weak shock is present ($\alpha=0.01$). In (b) with $\alpha=0.02$,
the shock disappears. Out of two choices (arrowed curves), the
one passing through the inner sonic point is preferred due to
higher dissipation.}

\noindent Fig. 1(a-b), adapted from Chakrabarti (1990)
which solves the fully viscous, isothermal
flows shows this behaviour. This prediction is verified by numerical
simulation of viscous flows$^{12}$. Fig. 2(a-b) shows the time variation
(curves drawn at intervals of $1000GM/c^3$)
of the Mach number and the angular momentum distribution in a
viscous ($\alpha=0.1$) flow which is sub-Keplerian at the outer boundary.
Here the angular momentum
is transported more efficiently in the post-shock flow and increasingly
higher centrifugal barrier pushes the shock further out making it weaker
in the process (2a) and making the disk shock free and quasi-Keplerian (2b).
These highly viscous, shock-free, quasi-Keplerian solutions self-consistently
pass through the inner sonic point. If the viscosity is very small,
a weak shock can survive$^{10,12}$ and the angular momentum distribution
remains sub-Keplerian everywhere except close to the marginally stable
orbit where it is super-Keplerian.
The dotted curve in Fig. 2a is the shock in the inviscid flow.

\vspace {6.10cm}
{\noindent \small Fig. 2(a-b) Time evolution of Mach number (a) and angular
momentum (b) variation as a function of logarithmic radial distance in a
viscous flow.}

\vspace {5.1cm}
{\noindent \small Fig. 3: Composite disk model around a black hole with
sub-Keplerian outer boundary condition and height dependent viscosity.}

We thus notice that one may have a thick, thin, slim or transonic (with or
without shocks) disk from an initially sub-Keplerian inflow at large distance
depending upon accretion rate and viscosity. Based on this experience,
in order to explain observations across the electromagnetic spectrum,
a generic model for the accretion disk could be built (Fig. 3).
A disk of this kind would form if there is a significant variation of
viscosity in the vertical direction inside the disk. Higher viscosity on the
equatorial plane produces an optically thick standard Keplerian disk, which is
vertically flanked by warm, optically thin halo of low angular momentum gas.
The halo forms a standing shock close to the black hole ($\sim 10\ R_g$).
The post-shock flow ($T_p \sim 10^{11}\ K$) heats up soft-photons
coming from the disk to produce observed $\gamma$-rays.
Such a model has the potential to explain most of the
steady state as well as time dependent behaviour of the continuum
and line emissions and is under active study$^{13}$.

In active galaxies, jets are also seen. It is generally assumed that
in the absence of a binary companion (which is a sink of angular momentum
in a binary system), jets has to carry away angular momentum of the disk.
Various jet models are constructed in order to achieve this goal. It is to
be remembered that the `angular momentum' problem is present only in models
which start with Keplerian flows at the outer boundary. In the sub-Keplerian
paradigm such problems are not present. Even in the presence of high viscosity,
the flow may have just enough (not excess!) {\it total} angular momentum
to redistribute itself to form a quasi-Keplerian disk. In the case of low
angular momentum flow, shear is strong only close to the black hole.
Strong shear produces very strong toroidal field. In the gas pressure dominated
hot disks (with $T_p> 4 \times 10^{10}\ K$) magnetic tension catastrophically
brings the flux tubes close to the black hole axis$^{14}$, ejecting
matter from the inner accretion disk in the form of jets. Thus the
jets would be blobby, rather than continuous. This general
feature may have been observed in detail in GRS 1915+105$^{15}$.\\

\centerline{REFERENCES}

\ni 1. Shakura, N.I. \& R.A. Sunyaev. 1973. Astron. Astrophys. {\bf 24:} 337.\\
\ni 2. Novikov, I. \& K.S. Thorne. 1973. in: Black Holes,
eds. C. DeWitt and B. DeWitt (Gordon and Breach, New York).\\
\ni 3. Sun, W.H. \& M.A. Malkan. 1989. Astrophys. J. {\bf 346:} 68.\\
\ni 4. Clavel, J. et al. 1990. Mon. Not. R. Astron. Soc. {\bf 246:} 668.\\
\ni 5. Chakrabarti, S.K. \& P.J. Wiita. 1994. Astrophys. J. {\bf 434:} 518.\\
\ni 6. Chakrabarti, S.K. 1995. Astrophys. J. (March 10th, in press)\\
\ni 7. Chakrabarti, S.K. 1995. Physics Reports (to appear).\\
\ni 8. Chakrabarti, S.K. 1990. Theory of Transonic Astrophysical Flows (World
Scientific, Singapore, 1990)\\
\ni 9. Chakrabarti, S.K. 1989. Astrophys. J. {\bf 347:} 365.\\
\ni 10. Chakrabarti, S.K. 1990. Mon. Not. R. Astron. Soc. {\bf 243:} 610.\\
\ni 11. Molteni, D., G. Gerardi \& S.K. Chakrabarti. 1994. Astrophys. J.
{\bf 436:} 249.\\
\ni 12. Chakrabarti, S.K. \& D. Molteni. 1995.  Mon. Not. R. Astron. Soc.
{\bf 272:} 80.\\
\ni 13.  Chakrabarti, S.K., L. Titarchuk, \& D. Kazanas. 1995.
(in preparation)\\
\ni 14. Chakrabarti, S.K. \& S. D'Silva. 1994. Astrophys. J. {\bf 424:} 138.\\
\ni 15. Mirabel, I.F. \& Rodriguez, L.F. 1994. Nature {\bf 371:} 46.\\
\end{document}